\documentclass[smallcondensed]{svjour3}            

\smartqed                          

\usepackage{graphicx}

\begin{document}

\title{Cold compression of nuclei induced by antiprotons}

\author{I.N. Mishustin \and A.B. Larionov}

\institute{ I.N. Mishustin \and A.B. Larionov \at 
            Frankfurt Institute for Advanced Studies, J.W. Goethe-Universit\"at,
            D-60438 Frankfurt am Main, Germany,
            and Russian Research Center Kurchatov Institute, 
            123182 Moscow, Russia\\
            \email{mishustin@fias.uni-frankfurt.de}  }


\maketitle

\begin{abstract}
On the basis of a dynamical Relativistic Mean Field (RMF) model
we study the response of a nucleus on the antiproton implanted
in its interior. We solve the Vlasov equation for the 
$\bar p$-nuclear system and show assuming a moderately attractive 
$\bar p$ optical potential that the compressed state is formed 
on a rather short time scale of about $4\div10$ fm/c. 
The evolution of the system after $\bar p$ annihilation is simulated
using the Giessen Boltzmann-Uehling-Uhlenbeck (GiBUU) transport model.
Finally, several sensitive observables to the $\bar p$ annihilation in a compressed 
nuclear configuration are proposed, e.g. the nucleon kinetic energy spectra 
and the total invariant mass distributions of produced mesons.
\end{abstract}

\section{Introduction}
\label{intro}

Creation of the compressed nuclear matter in laboratory is a hot topic
in the heavy ion community for more than three decades. The main interest 
here is to extract the information about the Equation of State of nuclear matter 
at high densities relevant for neutron star interiors. A well established way 
to produce compressed nuclear matter is to collide heavy ions with high energy.
However, in this case the produced system will be not only compressed, but also 
quite strongly excited thermally.

Recently several authors \cite{Tanida01,AY02,Buer02,Mish05,LMSG08}
discussed the possibility to compress the nuclear 
system without much thermal excitation by implanting in it a hadron 
($~\Lambda,~\bar K,~\bar p,~\bar\Lambda$) which has an attractive nuclear 
potential. It is believed that the antibaryon potentials are strongly 
attractive. The $G$-parity transformed proton potential is extremely deep: 
$U_{\bar p} \simeq -700$ MeV, while the phenomenological antiproton optical 
potential is much smaller $U_{\bar p} \simeq -(100\div350)$ MeV 
\cite{Wong84,Teis94,FGM05}. The static RMF calculations \cite{Buer02,Mish05} 
predict a strong nuclear compression effect by $\bar p$ and $\bar\Lambda$
even in the case of realistic potentials. For instance, the central nucleon 
density of the $^{16}$O core in the ground state $\bar p~\otimes ^{16}$O 
system is $(2\div4)\rho_0$, where $\rho_0=0.148$ fm$^{-3}$ is the normal
nuclear matter density.   

A natural question which arises here is whether the early annihilation 
of an antibaryon will kill the effect of compression? A simple estimate
of the stopped $\bar p$ life time in nuclear matter at the normal density
using vacuum annihilation cross section gives $\tau \simeq 2$ fm/c.
The phase space volume available in the baryon-antibaryon annihilation
could be, however, essentially reduced due to smaller effective masses
of the annihilating pair in nuclear matter. This effect can delay 
the annihilation up to $\sim 20$ fm/c \cite{Mish05}.  

In this talk we will present our results of dynamical simulations
of $\bar p$-nuclear systems to demonstrate the compression dynamics
and also to propose observable signals of possible $\bar p$
annihilation in a compressed nuclear environment. Sect.~\ref{model}
contains the short description of a model. The numerical results
are collected in Sect.\ref{results}, while the conclusions are given
in Sect.~\ref{conclusions}.

\section{The model}
\label{model}

Here we give only a very brief review of our dynamical model. 
The detailed description of the GiBUU model applied in calculations and 
of its RMF extension can be found in refs. \cite{LMSG08,GiBUU} and in 
refs. therein. 

In calculations we use the RMF Lagrangian of Ref. \cite{Mish05}, which includes 
the nucleon and antinucleon fields interacting with
the isoscalar-scalar $\sigma$-meson field and with the isoscalar-vector
$\omega$-meson field. The NL3 set of parameters was used.
This set provides a good description of nuclear matter and 
ground states of finite nuclei.   

The antinucleon couplings with $\sigma$- and $\omega$-fields were obtained by
rescaling the nucleon coupling constants as $g_{\sigma \bar N} = \xi g_{\sigma N}$
and $g_{\omega \bar N} = -\xi g_{\omega N}$, where $0 < \xi \leq 1$ is a scaling 
factor. Exact $G$-parity transformed antinucleon scalar and vector fields are obtained by 
the choice $\xi=1$. The antinucleon potentials consistent with the phenomenological 
ones are represented here by choosing $\xi=0.3$.

The dynamics of interacting nucleons and an antiproton was described within the transport
GiBUU model dealing with a coupled set of semiclassical kinetic equations (see \cite{LMSG08}
and refs. therein). The test particle technique in a parallel ensemble mode has been used to 
solve these equations.

An antiproton was initialized according to the gaussian density distributions in 
coordinate and momentum space. The centroids of the gaussians were set to describe
the particle at the center of a nucleus (${\bf r}=0$) at rest (${\bf p}=0$).
After initialization the system was left to evolve according to the kinetic
equations without collision terms until $\bar p$ annihilation at some preselected 
time moment $t_{\rm ann}$. At this time a sudden annihilation of the antiproton with 
the closest nucleon was simulated on the basis of the string model 
JETSET. Once the annihilation is simulated, further evolution of the system of 
nucleons and mesons has been followed by solving the full kinetic equations including 
the collision terms. In this way, we have taken into account, in-particular, important 
processes of pion rescattering and absorption: $\pi N \to \Delta \to \pi N$ or 
$\pi N \to \Delta$, $\Delta N \to N N$.

\section{Numerical results}
\label{results}

\begin{figure}
\begin{center}
   \includegraphics[bb = 64 50 570 705, scale = 0.5]{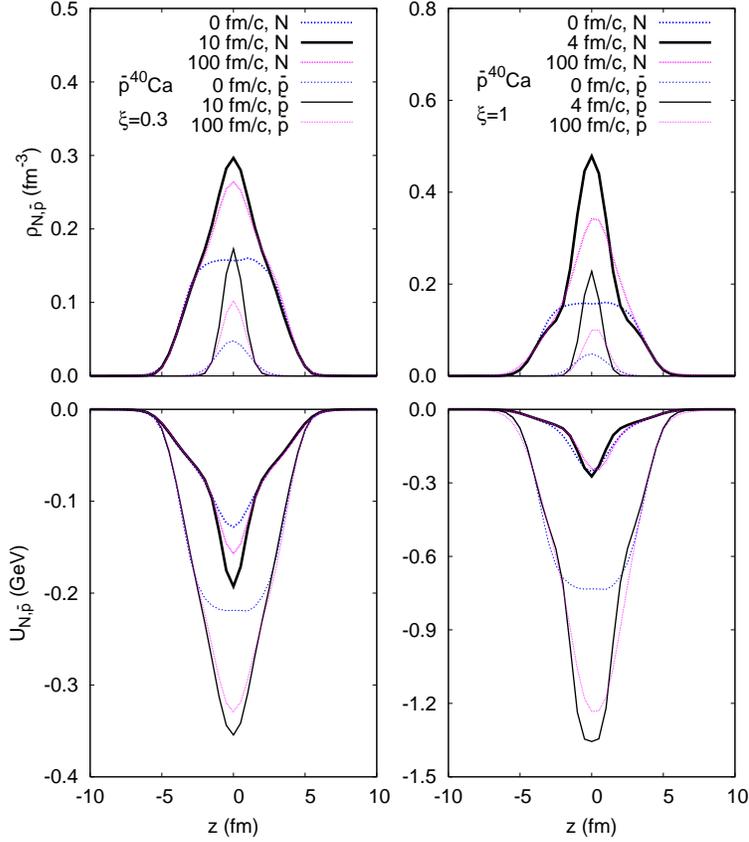}
\end{center}
\caption{\label{fig:rho_U_vs_z} Nucleon and antiproton densities
(top panels) and potentials (bottom panels) vs coordinate $z$ on the axis passing
through the center of the $\bar p$$^{40}$Ca system at selected times indicated
in the figure.
The calculations with the scaling factor $\xi=0.3$ ($\xi=1$) are shown in the left (right)
panels.}
\end{figure}

First, we have studied the compression dynamics of a nucleus under the influence 
of an antiproton implanted in its center, completely neglecting the 
collision terms in kinetic equations, i.e. by taking into account the
mean field only. The upper panels of Fig.~\ref{fig:rho_U_vs_z} show the 
axial density distributions for nucleons and an antiproton at different 
times. In the lower panels of Fig.~\ref{fig:rho_U_vs_z}, the
potentials defined as $U_j = g_{\omega j} \omega^0 + g_{\sigma j} \sigma$
($j=N,~\bar p$) are also shown.
The nucleon density at the center grows quickly reaching its maximum
already at $t=10$ fm/c for the reduced $\bar p$-couplings and even faster
at, $t=4$ fm/c, for the $G$-parity-motivated couplings. The maximum nucleon
density reached in this process is $\rho_N^{\rm max} \simeq 2\rho_0$
($\rho_N^{\rm max} \simeq 3\rho_0$) for $\xi=0.3$ ($\xi=1$).
The compression of the nuclear system leads also to the growth 
of the antiproton density at the center. The system stabilizes on
the time scale of several tens fm/c.

\begin{figure}
\begin{center}
   \includegraphics[bb = 30 90 537 750, scale = 0.5]{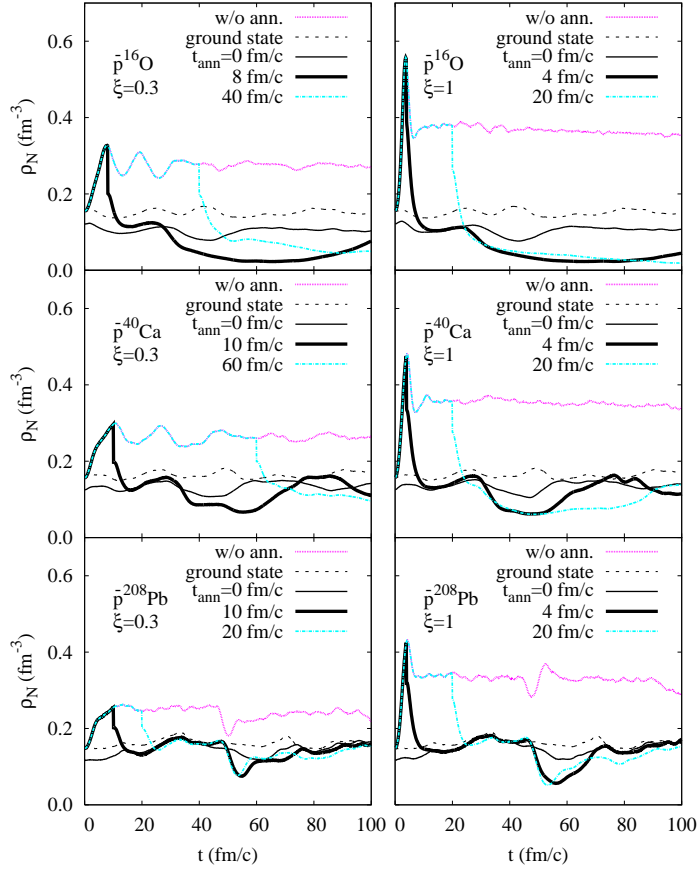}
\end{center}
\caption{\label{fig:rho_vs_t} Time evolution of the nucleon densities
at the centers of the $\bar p$$^{16}$O, $\bar p$$^{40}$Ca and $\bar p$$^{208}$Pb 
systems for two values of the scaling factor $\xi=0.3$ (left panels) and $\xi=1$ 
(right panels). The dotted line shows 
the calculation without $\bar p$ annihilation. The thin solid, thick solid and dash-dotted 
lines show the results with annihilation simulated at various times $t_{\rm ann}$
indicated in the figure. The dashed lines show the central nucleon density for 
the corresponding ground state nucleus without an antiproton.}
\end{figure}

Compression process is better demonstrated in Fig.~\ref{fig:rho_vs_t}, where
the central nucleon density evolution is presented for the  three systems:
$\bar p$$^{16}$O, $\bar p$$^{40}$Ca and $\bar p$$^{208}$Pb. For all the systems
a compressed state is reached within $t=4\div10$ fm/c and exists up to
$t=100$ fm/c. At later times the stability is gradually lost due to the test particle 
escape beyond the computational grid. 

To understand how compression can possibly influence observables, we have simulated
the annihilation of an antiproton at different preselected times $t_{\rm ann}$.
The evolution of a central nucleon density for different $t_{\rm ann}$ is also
shown in Fig.~\ref{fig:rho_vs_t}. If the annihilation takes place in a compressed
state, the system expands and, after $30\div40$ fm/c, the central density drops 
below $\rho_0$. For the light and medium systems, $\bar p$$^{16}$O and 
$\bar p$$^{40}$Ca, the central nucleon density stays below $0.6 \rho_0$ for about 
$30\div50$ fm/c, which is sufficient for the small density perturbations to grow 
according to the spinodal mechanism \cite{CCR04}. This process should eventually 
lead to the multiple fragment formation.

\begin{figure}
\begin{center}
   \includegraphics[bb = 120 75 626 730, scale = 0.5]{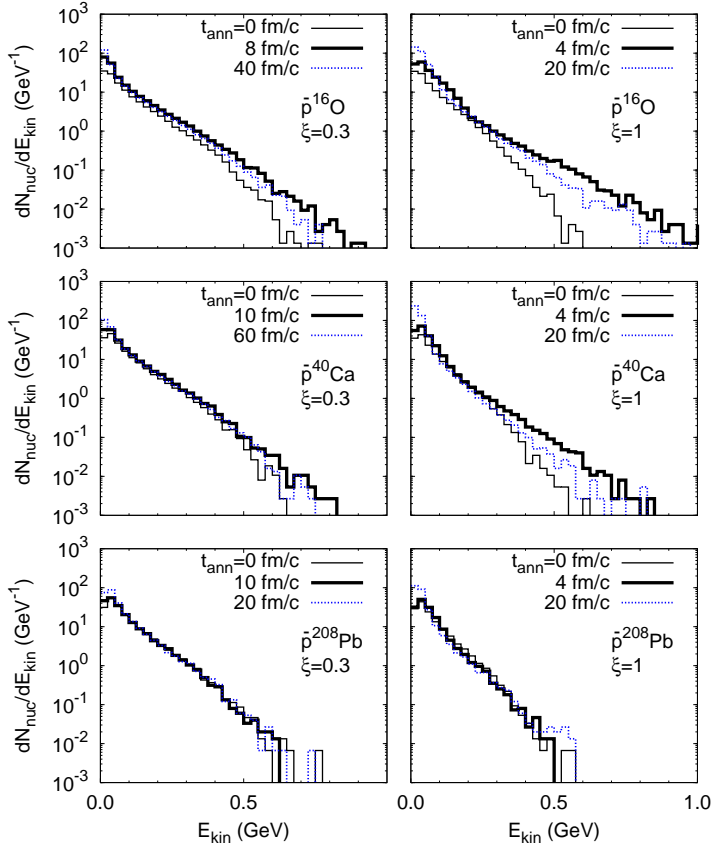}
\end{center}
\caption{\label{fig:Ekin_spectra} Kinetic energy spectra of 
emitted nucleons in the c.m. frame for various $\bar p$A systems and values of the parameter
$\xi$ as in Fig.~\ref{fig:rho_vs_t}. Different histograms correspond to different values of 
the annihilation time $t_{\rm ann}$ indicated in the key.}
\end{figure}

Fig.~\ref{fig:Ekin_spectra} shows the kinetic energy spectra of nucleons emitted
from the system after $\bar p$-annihilation. They are mostly kicked-out
by pion rescattering via intermediate $\Delta(1232)$ resonance:
$\pi N \to \Delta \to \pi N$ \cite{Cahay82}. In the case of annihilation in the 
compressed state, the nucleon spectrum acquires a high-energy tail with clearly 
stronger effect for lighter systems. This can be explained by the following 
mechanism: During the compression stage the nucleons in the central zone increase 
their kinetic energies due to falling down into the deep potential well created by 
the antiproton (see lower panels of Fig.~\ref{fig:rho_U_vs_z}). After 
$\bar p$-annihilation the potential well disappears suddenly, and the fast nucleons
get released. This mechanism combined with additional kicks by pions
leads to a visible enhancement of the energetic nucleon production.  

\begin{figure}
\begin{center}
   \includegraphics[bb = 64 75 590 730, scale = 0.5]{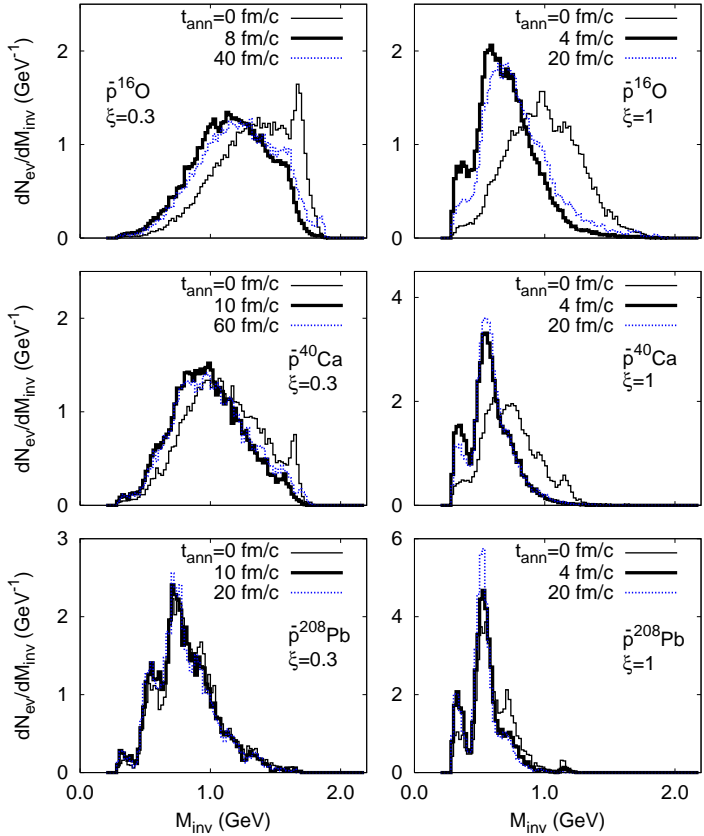}
\end{center}
\caption{\label{fig:Minv_spectra} Distributions of annihilation events in 
the total invariant mass of emitted mesons for various $\bar p$A systems and values of 
the parameter $\xi$ as in Fig.~\ref{fig:rho_vs_t}. Different histograms correspond to 
different values of the annihilation time $t_{\rm ann}$ indicated in the key.}
\end{figure}

In Fig.~\ref{fig:Minv_spectra} we show the distribution of annihilation events
in the total invariant mass $M_{\rm inv}$ of emitted mesons. In the absence of any 
in-medium effects, this distribution should be sharply peaked at $2m_N$. The final 
state interactions of mesons, i.e. rescattering and absorption on nucleons, lead
to the strong spreading of the distribution toward smaller $M_{\rm inv}$. Additionally,
the mean-field effects reduce the invariant energy $\sqrt{s}$ of annihilating 
$\bar p$-nucleon pair as $\sqrt{s} \simeq 2m_N + U_N + U_{\bar p} < 2m_N$. This further 
shifts the distribution to smaller $M_{\rm inv}$. The effect is especially strong for 
annihilations in compressed configurations.   

The above results have been obtained for the artificial case of an antiproton sitting
at rest in the center of a nucleus. It is, certainly, more realistic to study the 
compression of the nucleus induced by moving $\bar p$. As an example, we have considered 
the central $\bar p$$^{12}$C collision at $E_{\rm lab}=180$ MeV. In order to compute the 
mean field, we have performed simulations for 1000 parallel ensembles of test particles.  
The probability of $\bar p$-annihilation within the time interval $\Delta t$ was
calculated as $P_{\rm ann} = 1 - \exp(-\Gamma_{\bar p}^{\rm ann}\Delta t)$, where 
$\Gamma_{\bar p}^{\rm ann}$ is the annihilation width and $\Delta t=0.2$ fm/c is the time 
step of the simulation. In each run, the annihilation time for all parallel ensembles
has been chosen randomly using $P_{\rm ann}$. Further details of the model implementation 
for the $\bar p$-induced reactions on nuclei will be published elsewhere. 

\begin{figure}
\begin{center}
   \includegraphics[bb = 64 75 590 730, scale = 0.5]{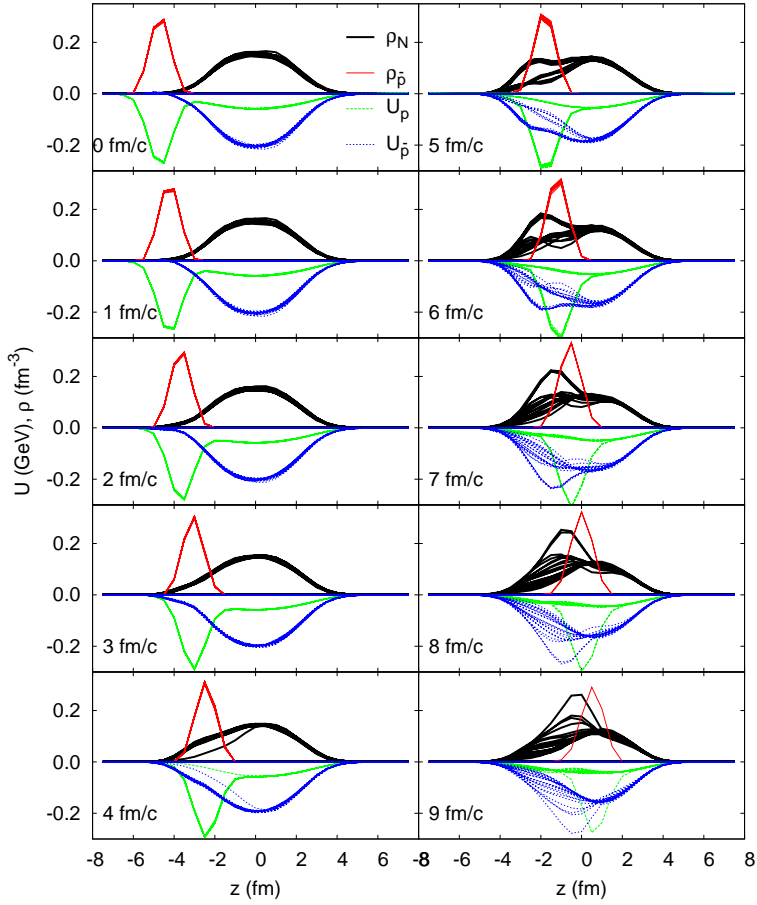}
\end{center}
\caption{\label{fig:rhoz_pbarC12_180MeV_coh} Nucleon and antiproton density 
and potential profiles along collision axis at different time moments for the
$\bar p$(180 MeV)+$^{12}$C collision at the impact parameter $b=0.5$ fm. The mean field 
was computed with reduced $\bar p$ couplings ($\xi=0.3$).}
\end{figure}

In Fig.~\ref{fig:rhoz_pbarC12_180MeV_coh} we show the density distributions 
of nucleons and of the antiproton along the beam axis for 20 computed runs. 
One can see that in 19 runs the antiproton annihilation happened at the periphery 
of the nucleus. And only in a single run the antiproton has safely reached the 
central zone. In this run, the nucleon density has been increased up to $\simeq 2\rho_0$ 
by moving $\bar p$. This means that with $\sim 5\%$ probability one can expect the
compression effect even in realistic central $\bar p$-nucleus collisions.

\section{Conclusions}
\label{conclusions}

To summarise, we have performed the dynamical RMF calculations of the cold compression
of nuclei by an antiproton implanted in their centers. The calculations in a pure
mean field mode resulted in a fast increase of the central nucleon density up to 
$2\div3\rho_0$ within the time interval of $4\div10$ fm/c. Without annihilation,
the system stabilizes in a compressed state, which has been numerically traced 
until 100 fm/c. The density profiles of compressed $\bar p$-nuclear systems are close 
to the ones obtained earlier by static RMF calculations \cite{Buer02,Mish05}.

Due to in-medium effects, the $\bar p$ annihilation can be suppressed.
In this case the time needed for a compression can be comparable with the 
life time of an antiproton with respect to the annihilation. Therefore, the 
compression can, indeed, manifest itself in some $\bar p$-nucleus 
collision events before the antiproton annihilation.

It has been demonstrated, that the events with nuclear compression
have a probability of few percent even in $\bar p$(180 MeV)+$^{12}$C 
central collisions. At higher energies, the fraction of the compression events 
is expected to be higher due to smaller $\bar p$-nucleon annihilation cross section.

In the present calculations we have used quite moderate values for the
attractive real part of $\bar p$ optical potential. Actual determination 
of $U_{\bar p}$ requires a very careful comparison with experimental data
on elastic and inelastic scattering and absorption of antiprotons on nuclei.
We propose to perform such measurements at the future FAIR facility in Darmstadt.
In this respect we find very promising to use the event-by-event transverse 
momentum correlations proposed recently to determine the $\bar\Lambda$ 
optical potential \cite{Pochodzalla}. We believe that the same method can also be 
applied for the determination of the $\bar p$ optical potential.   

\begin{acknowledgements}
We thank L.M. Satarov, W. Greiner, H. St\"ocker, Th.J. B\"urvenich 
and I.A. Pshenichnov for productive discussions. 
This work was supported in part by the DFG Grant 436 RUS 113/711/0-2
(Germany) and the Grant NS-3004.2008.2 (Russia).
\end{acknowledgements}


\begin{thebibliography}{99}

\bibitem{Tanida01} Tanida, K. et al.: 
Measurement of the B(E2) of $^7_\Lambda$Li and Shrinkage 
of the Hypernuclear Size.
Phys. Rev. Lett. {\bf 86}, 1982-1985 (2001)

\bibitem{AY02} Akaishi, Y. and Yamazaki, T.:
Nuclear $\bar{\rm K}$ bound states in light nuclei.
Phys. Rev. C {\bf 65}, 044005 1-9 (2002) 

\bibitem{Buer02} B\"urvenich, T., Mishustin, I.N.,
Satarov, L.M., Maruhn, J.A., St\"ocker, H., and Greiner, W.:
Enhanced binding and cold compression of nuclei due to admixture 
of antibaryons. 
Phys. Lett. B {\bf 542}, 261-267 (2002) 

\bibitem{Mish05} Mishustin, I.N., Satarov, L.M., B\"urvenich, T.J.,
St\"ocker, H., and Greiner, W.:
Antibaryons bound in nuclei.
Phys. Rev. C {\bf 71}, 035201 1-32 (2005)

\bibitem{LMSG08} Larionov, A.B., Mishustin, I.N., Satarov, L.M.,
and Greiner, W.:
Dynamical simulation of bound antiproton-nuclear systems 
and observable signals of cold nuclear compression.
Phys. Rev. C {\bf 78}, 014604 1-14 (2008)

\bibitem{Wong84} Wong, C.-Y., Kerman, A.K., Satchler, G.R.,
MacKellar, A.D.:
Ambiguity in antiproton-nucleus potentials from antiprotonic-atom data.
Phys. Rev. C {\bf 29}, 574-580 (1984)

\bibitem{Teis94} Teis, S., Cassing, W., Maruyama, T.,
and Mosel, U.:
Analysis of subthreshold antiproton production in p-nucleus and 
nucleus-nucleus collisions in the relativistic Boltzmann-Uehling-Uhlenbeck 
approach.
Phys. Rev. C {\bf 50}, 388-405 (1994)

\bibitem{FGM05} Friedman, E., Gal, A., Mare\v{s}, J.:
Antiproton-nucleus potentials from global fits to antiprotonic 
X-rays and radiochemical data.
Nucl. Phys. A 761, 283-295 (2005)

\bibitem{GiBUU} http://theorie.physik.uni-giessen.de/GiBUU

\bibitem{CCR04} Chomaz, Ph., Colonna, M., and Randrup, J.:
Nuclear spinodal fragmentation.
Phys. Rep. {\bf 389}, 263-440 (2004) 

\bibitem{Cahay82} Cahay, M., Cugnon, J., Jasselette, P. and Vandermeulen, J.:
Antiproton annihilation inside nuclei.
Phys. Lett. B {\bf 115}, 7-10 (1982)

\bibitem{Pochodzalla} Pochodzalla, J.:
Exploring the potential of antihyperons in nuclei with antiprotons.
arXiv:0807.3302v2 and contribution to these proceedings

\end{thebibliography}
\end{document}